\documentclass[twocolumn]{aastex631}

\usepackage{tabularx}
\usepackage{amsmath}

\newcommand\p{\partial}
\newcommand\pd[2]{ \frac{\p #2}{\p #1} }
\newcommand\eps{\epsilon_0}

\shorttitle{Electrostatic Bursts}
\shortauthors{Afify et al.}

\begin{document}

\title{Electrostatic Bursts Generated by the Ion-Ion Acoustic Instability
with Solar Wind Plasma Parameters}
\correspondingauthor{Mahmoud Saad Afify}
\email{Mahmoud.Ibrahim@ruhr-uni-bochum.de\\Mahmoud.Afify@fsc.bu.edu.eg}

\author[0000-0002-2088-0354]{Mahmoud Saad Afify}
\affiliation{Theoretische Physik I, Ruhr-Universität Bochum, Bochum, Germany}
\affiliation{Department of Physics, Faculty of Science, Benha University, Benha 13518, Egypt}
\author[0000-0002-6189-9158]{Jürgen Dreher}
\affiliation{Theoretische Physik I, Ruhr-Universität Bochum, Bochum, Germany}
\author[0000-0002-7644-7381]{Kevin Schoeffler}
\affiliation{Theoretische Physik I, Ruhr-Universität Bochum, Bochum, Germany}
\author[0000-0001-9293-174X]{Alfredo Micera}
\affiliation{Theoretische Physik I, Ruhr-Universität Bochum, Bochum, Germany}
\author[0000-0002-5782-0013]{Maria Elena Innocenti}
\affiliation{Theoretische Physik I, Ruhr-Universität Bochum, Bochum, Germany}

\begin{abstract}
  This study is motivated by recent observations from the Parker Solar Probe (PSP) mission, which have been identified as the ion-acoustic waves from 15 to 25 solar radii. These observations reveal characteristic sequences of narrow-band, high-frequency bursts exceeding 100 Hz embedded into a slower evolution around 1 Hz, persisting for several hours. To explore the potential role of the ion-acoustic instability (IAI) in these phenomena, we begin by reviewing classical findings on the IAI within the framework of linear kinetic theory. Focusing on proton distributions comprising both a core and a beam component, we analyze the IAI instability range and growth rates within the parameter regime relevant to PSP observations. Our findings indicate that the IAI can indeed occur in this regime, albeit requiring electron-to-core and beam-to-core temperature ratios slightly different from reported values during electrostatic burst detection. Furthermore, employing 1D kinetic plasma simulations, we validate the growth rates predicted by linear theory and observe the saturation behavior of the instability. The resultant nonlinear structures exhibit trapped proton beam populations and oscillatory signatures comparable to those observed, both in terms of time scales and amplitude.
\end{abstract}


\section{Introduction} \label{sec:intro}


The solar wind is an almost collisionless plasma consisting mainly of electrons and protons, where kinetic-scale processes abound~\citep{vstverak2008electron, Matteini2012, innocenti2020collisionless, micera2020particle, Sun2021, Micera2021, Verscharen2022, bandyopadhyay2022interplay}. In recent years, missions such as Parker Solar Probe~\citep{Fox2016} and Solar Orbiter \citep{muller2020solar}, together with a re-evaluation of observations from previous missions, such as Helios~\citep{rosenbauer1977survey}, sparked renewed interest in the kinetic physics of the solar wind and in its role in energy transport and dissipation at the small kinetic scales \citep{bruno2013solar, scime1994regulation, Perrone2022, Verscharen2019}.
One of the most striking kinetic features of solar wind velocity distribution functions (VDFs) is the ubiquitous presence of proton beams, which can originate from a number of different processes, including magnetic reconnection~\citep{Dai2021, Phan2022} and turbulence~\citep{Valentini2011, Perrone2011, Valentini2014}. During its first perihelion, PSP observed that proton beam VDFs often assume a ``hammerhead" shape during periods of coherent ion-kinetic waves ~\citep{Verniero2022, pezzini2024fully}. 

The presence of multiple proton populations can result in a number of instabilities. Sources of instability-driving free energy can in fact include both ion temperature anisotropy and the presence of drifting ion populations~\citep{GaryBook, Matteini2012, Ďurovcová2019, Klein2019, Klein2021, Mihailo2021, Mostafavi2022, Mihailo2023, Ofman2023}. Examples of instabilities driven by multiple particle populations include the current-driven instability, caused by the relative drift between electrons and ions; the ion-acoustic heat flux instability, caused by the relative drift between strahl and core electrons; and the ion-ion-acoustic instability, caused by the relative drift between two ion populations \citep{GaryBook}. 

In the electron-ion instability, a small fraction of resonant electrons drives the Landau resonance, causing the propagation of ion-acoustic waves. In the ion-ion acoustic instability, on the contrary, these waves are triggered by resonant ions.
These instabilities are very sensitive to the electron-to-ion temperature ratio, where $T_e/T_i \gg 1$ is required to avoid strong damping \citep{Lemons1979}. Other possible sources of the ion-acoustic instability \citep{Cairns1992, Saito2017} are the decay of Langmuir waves and nonlinear turbulence of whistler modes~\citep{Cairns1992, Saito2017}. 

Ion-acoustic waves, also known as kinetic slow modes \citep{Verscharen2013}, are electrostatic phenomena that have been observed in the solar wind for over 40 years \citep{Gurnett1977, Kurth1979, Gurnett1991, Pisa2021, Graham2021} and are ubiquitous in collisionless shocks \citep{Wilson2007}. At 55 solar radii, PSP detected electrostatic waves that exhibited a wide bandwidth and short duration ~\citep{Mozer2020}. These waves were identified as Doppler-shifted ion-acoustic waves because their wave vector is aligned with the background magnetic field and they are linearly polarized. The proton VDF observed in coincidence with these waves consisted of a core and a beam and did not show evidence of ion heating as a consequence of the observed activity.
PSP also reported a variety of the ion-acoustic modes at a distance of 35 solar radii around the boundaries of magnetic field switchbacks~\citep{neugebauer2013double, Wilson2018, Mozer2020,tenerani2021evolution}. In a nonlinear stage, PSP measurements suggest that the ion-acoustic modes can result in ion and electron holes \citep{Mozer2021}. Recently, \citet{Malaspina2024} reported PSP observations of IAWs sunward of $\sim 60$ solar radii and identified their driver as cold, impulsively accelerated proton beams near the ambient proton thermal speed.

The motivation behind this work is recent observations of ion-acoustic waves recorded by PSP from 15 to 25 solar radii \citep{Mozer2021a}, that are different from those seen in previous measurements. They occur as short-lived, narrow-banded electrostatic bursts with frequencies close to one hundred Hertz embedded in envelopes exhibiting much lower frequencies, of the order of a few Hertz \citep{Mozer2023a}, resulting in a very regular
pattern that can persist for several hours. Their phase velocities are estimated to roughly match the local ion-acoustic speed. Given that electrons are significantly hotter than the core ions during this wave activity and that proton beams are ubiquitously observed in the solar wind, \citet{Mozer2020} supposed that in the presence of two resonant counter-steaming ion populations, the ion-acoustic instability could be triggered, giving rise to the occurrence of these ion-acoustic bursts. The measured proton distribution functions have a plateau during the wave activity \citep{Mozer2021a} and can be modeled as a core-beam distribution consisting of two drifting Maxwellians.

In this study, we concentrate on the possible mechanism underlying the reported observations of ion-acoustic modes from 15 to 25 solar radii by \citet{Mozer2021a}. We intend to understand if the ion-ion-acoustic instability can arise under the solar wind parameters reported there. To do that, in Section~\ref{sec:theory}, we solve the electrostatic kinetic dispersion relation for the IAI, and we identify parameter ranges of stability and instability, which we compare with the parameters reported by \citet{Mozer2021a}.
Then, in Section~\ref{sec:simulation}, we run 1D-1V Vlasov kinetic simulations using parameters that roughly fall into the regime reported from the observations and demonstrate the occurrence of the IAI.
We use these simulations not only to validate the theoretical predictions for the growth rate and wavenumber of the ensuing instability, but also to explore the nonlinear phase of the instability, which results in the generation of characteristic electrostatic signatures that compare well with those observed.
Finally, in the Conclusion, we review our results. We focus on similarities and differences between the observations and our simulations, and we illustrate our plans for future work.

\section{Plasma Model and Linear Theory} \label{sec:theory}

Following \citet{Gary1987} and \citet{GaryBook}, we describe the
plasma as a collection of three components, I) a proton core
distribution, II) a proton beam, and III) electrons, which we label with indices
$c$, $b$, and $e$, respectively. Assuming a one-dimensional setting
with relevant space coordinate $x$ and corresponding velocity
component $v$, each of the components $\alpha = c, b, e$ is described
by a time-dependent distribution function $f_\alpha(x, v, t)$ that
obeys Vlasov's equation for collisionless electrostatic interaction,
\begin{equation}
\pd{t}{f_\alpha} + v \pd{x}{f_\alpha} + \frac{q_\alpha}{m_\alpha} E \pd{v}{f_\alpha} = 0. \label{eq:vlasov}
\end{equation}
Here, the  $x$-component of the electric field, $E(x, t)$, couples the components via Gauss' law,
\begin{equation}
 \pd{x}{E} = \frac{1}{\eps} \sum_\alpha q_\alpha n_\alpha
 = \frac{1}{\eps} \sum_\alpha q_\alpha \int f_\alpha \, dv
 \label{eq:gauss}
 \end{equation}
where $\eps$ denotes the vacuum permittivity, $m_\alpha$ and $q_\alpha$ are the particles'
masses and charges, and $n_\alpha(x, t)$ are the particle number densities, respectively.
According to standard Landau theory, linearizing this system around a homogeneous equilibrium configuration with
distributions $f_{0, \alpha}(v)$ and vanishing electric field, and applying a perturbation ansatz $\sim e^{i(kx - \omega t)}$ results
in the dispersion relation
\begin{equation}
k^2 = \sum_{\alpha} \frac{\omega_{p, \alpha}^2 }{n_\alpha} \int\frac{dv}{v - \omega/k} \frac{\p f_{0, \alpha}}{\p v} \label{eq:disprela}
\end{equation}
with plasma frequencies $\omega_{p, \alpha} = \sqrt{ n_\alpha q_\alpha^2 / \eps m_\alpha }$.

In this work, we will focus on drifting Maxwellian distributions
\begin{equation}
f_{0, \alpha}(v) = \frac{n_\alpha}{\sqrt{ 2\pi} v_{th, \alpha} } e^{-(v - V_{\alpha})^2 / (2 v_{th, \alpha}^2 )} \label{eq:maxwellian}
\end{equation}
with drift velocities $V_{\alpha}$ and  thermal velocities $v_{th, \alpha} = \sqrt{T_\alpha / m_\alpha}$ (the temperatures $T_\alpha$ are given in
energy units), for which the dispersion relation can be expressed in terms of the plasma dispersion function.
Furthermore, we will assume  $m_c = m_b = 1836\;m_e $ and charges $ q_c = q_b = - q_e $, suitable for proton/electron plasma,
set $V_c = 0$ (i.e., choose the core ion distribution as a reference frame), and assume the system to be neutral, $n_e = n_c + n_b$,
and current-free, $V_e = -n_b V_b / n_e $.
The core thermal velocity $v_{th, c}$, plasma frequency $\omega_{p, c}$, and Debye length $\lambda_{D, c} = V_{th, c} / \omega_{p, c}$ serve
as normalization values for speed, time, and length, respectively.
The remaining key parameters of this setup are the density ratio between the beam and core, $n_b/n_c$, the beam-core drift velocity, $V_D=V_b-V_c$, and the temperature ratios between electrons and core ions, $T_e/T_c$, and beam and core ions, $T_b/T_c$.

The dispersion relation (\ref{eq:disprela}), (\ref{eq:maxwellian}) has been solved for parameters motivated by the PSP observations reported in~\citet{Mozer2021a}, using standard root-finding methods. For convenience, we tabulate the relevant observational parameters in Table 1, which also contains the crucial ratios between beam and core density, beam and core temperature, and electron and core temperature, respectively, which were calculated using the parallel temperatures as they are the relevant ones for the IAI to develop parallel to the magnetic field. \\
\begin{table}[h]
\centering
\resizebox{\columnwidth}{!}{%
\begin{tabular}{lcc}
\hline \hline
\textbf{Table 1:} Observational parameter values as given by \citet{Mozer2021a} \\ 
\hline \hline
Ion core density ($n_c$) & 1220 & $\mathrm{cm}^{-3}$ \\
Ion beam density ($n_b$) & 31 & $\mathrm{cm}^{-3}$ \\
$n_b/n_c$ & 0.025 & \\
Relative drift between core and beam ($V_D$) & -180 & $\mathrm{km/s}$ \\
Perpendicular core temperature ($T_{c,\perp}$) & 10 & $\mathrm{eV}$ \\
Perpendicular beam temperature ($T_{b,\perp}$)& 17 & $\mathrm{eV}$ \\
Core anisotropy $T_{c,\perp } ~/ ~ T_{c,\| }$ & 1.3 & \\
Beam anisotropy $T_{b,\perp } ~/ ~ T_{b,\| }$ & 0.8 & \\
Electron temperature ($T_e$) & 50 & $\mathrm{eV}$ \\
$T_{e} ~/ ~ T_{c,\|}$ & 6.5 & \\
$T_{b,\|} ~/ ~T_{c,\|}$ & 2.75 & \\
$V_D ~/ ~\sqrt{T_{c,\|} /m_p}$ & 6.63 & \\
$\boldsymbol{\beta_{c,\|}}$ & \textbf{0.093} & \\
$\boldsymbol{\beta_{e,\|}}$ & \textbf{0.643} & \\
\textbf{Magnetic field} & \textbf{200} & $\mathrm{\textbf{nT}}$ \\ 
\hline \hline 
\end{tabular}
}
\end{table}

\begin{figure}[ht!]
\plotone{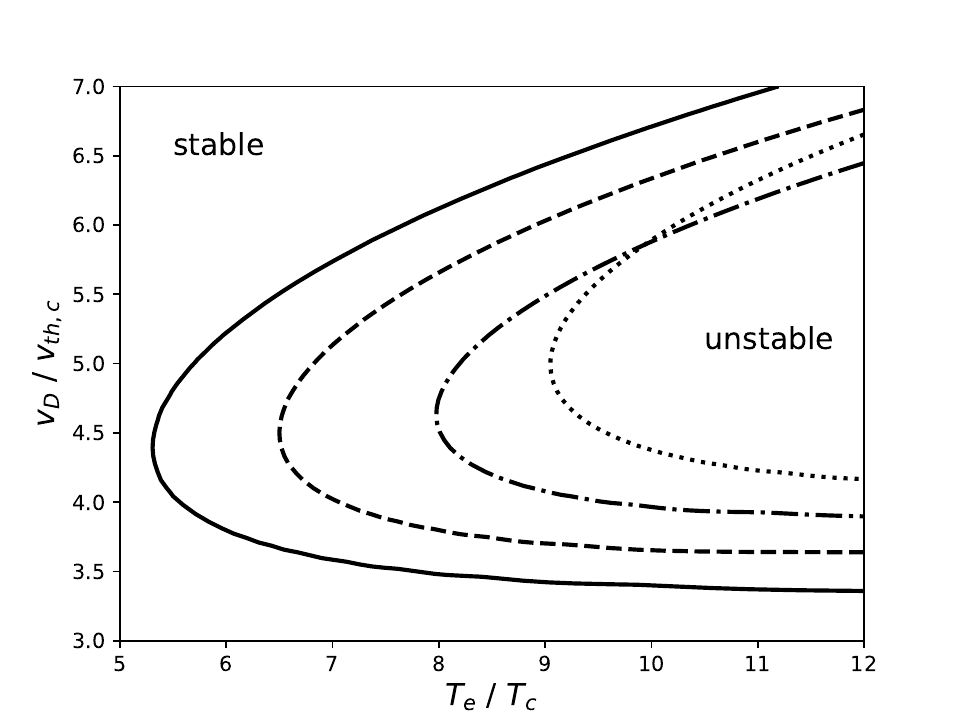}
\caption{
  Stability region of the ion-ion acoustic instability for $T_b = T_c$ with different values of
  $n_b / n_c = 0.1$ (solid), $0.05$ (dashed), and $0.025$ (dash-dotted), and for
  $T_b / T_c = 1.5$ with $n_b / n_c = 0.025$  (dotted).
  }
\label{fig:stability}
\end{figure}

Fig. \ref{fig:stability} displays the region of instability in the $T_e/T_c$ vs. $V_D/V_{th,c}$ plane, i.e. combination of these parameters for which a solution with positive
growth rate $\gamma = \mathrm{Im}( \omega)$ exists, using three different density ratios $n_b/n_c$.  
The parameter set of the solid line, i.e., $n_b = n_c/10$ and $T_b = T_c$, is already
addressed by \citet{Gary1987} and used here as a reference. It is clear from Fig. \ref{fig:stability} that for high enough $T_e/T_c$, a region of instability for the IAI exists, bounded by a minimum and maximum value for the core-to-beam drift velocity. The unstable region increases with an increasing $n_b/n_c$ ratio. Note that this matches the results from Fig. 1 of \citet{Gary1987}, while the upper limit on $V_D$ for instability onset is not present in their Fig. 6. Extending the study to cases with lower ratios of $ n_b / n_c $, i.e., towards values given in  \citep{Mozer2021a}, and also increasing the beam temperature above that of the core, we note that the necessary temperature ratio $ T_e / T_c $ for the existence of the IAI rapidly increases as a consequence of the smaller beam kinetic energy and the effect of resonant damping in the beam population. Specifically, for $ n_b / n_c = 0.025 $ and $ T_b = T_c $, the threshold for $ T_e $ is approximately $ 7.9 \times T_c $. 

From these considerations, it is clear that the values of Table 1 together with the modeling assumptions described above, predict stability with respect to the IAI. However, the observations are highly variable, and the model gives only an approximation to the actual setting. This, together with the fact that the occurrence of the observed electrostatic bursts appears to be triggered by changes in the environment and that the instability threshold seems to be highly sensitive to the crucial parameters, suggests that the observed parameters that could be measured during the saturation phase are close to the border of instability, which leads us to use slightly different parameter values to investigate the IAI in cases where the instability threshold is crossed. Specifically, we used $T_e/T_c=10$, $V_D/V_{th,c}=5$, and $T_b/T_c=1$ or $1.5$.      

\begin{figure}[ht!]
\plotone{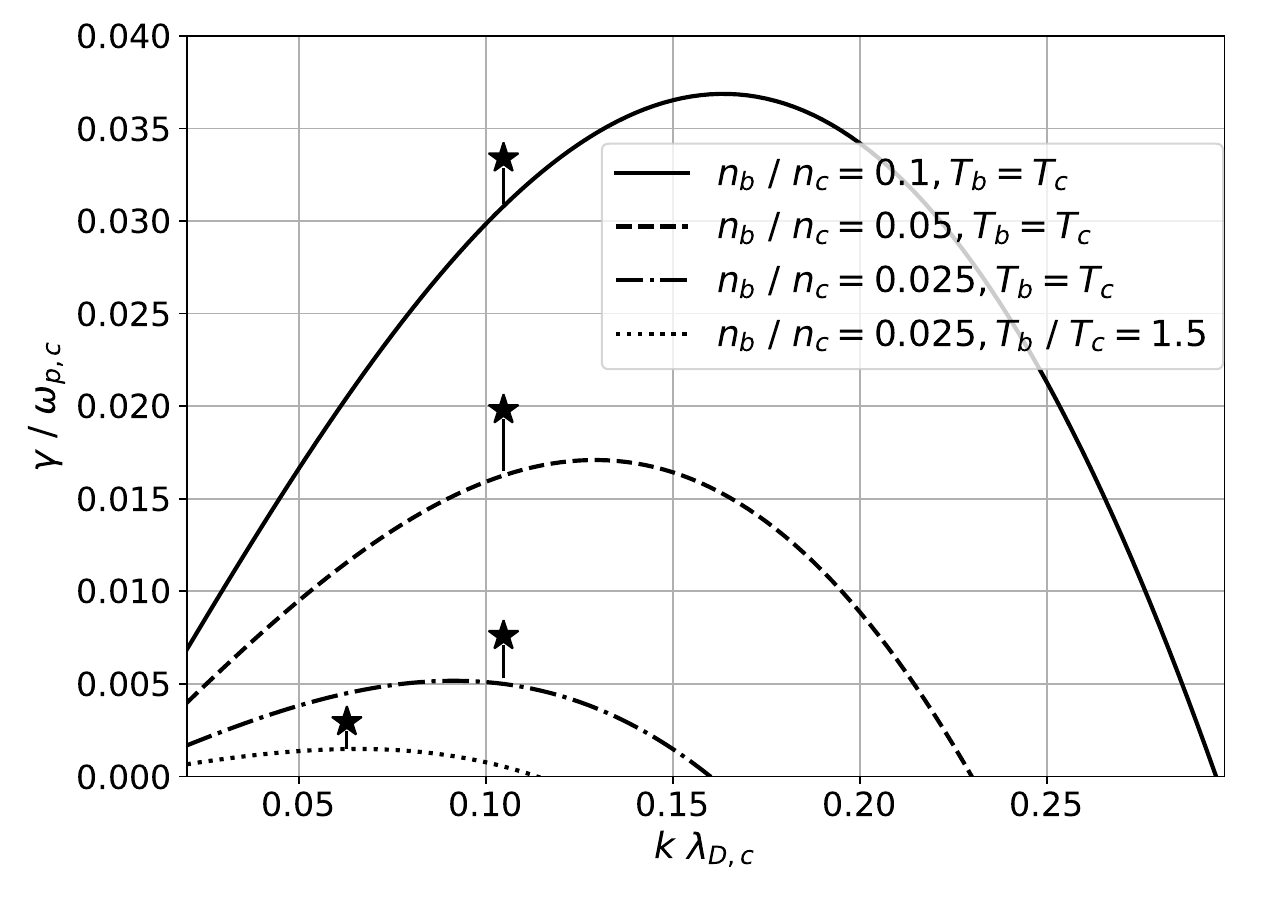}
\caption{
  Normalized growth rate $\gamma$ vs. wave number $k$ of the instability for $T_e / T_c = 10$ and $V_D / V_{th,c} = 5$
 with different beam-to-core ratios of density and temperature. Stars denote growth rates obtained from Fig. \ref{fig:growth} for the numerical simulation runs presented in Section \ref{sec:simulation}, and from a run of the reference case with $n_b / n_c = 0.1$.
}
\label{fig:disprel}
\end{figure}

Fig. \ref{fig:disprel} displays the growth rate with $ T_e/T_c = 10 $, for which all four beam-to-core density and temperature cases illustrated in Fig. \ref{fig:stability} are unstable. It demonstrates that smaller $n_b/n_c$ and larger $T_b/T_c$ both have a stabilizing influence, reflected
by rapidly diminishing growth rates, a reduced range of wavenumbers susceptible to instability, and a shift of the maximum growth rate towards longer wavelengths.
In particular, for the most dilute beam, the maximum growth is predicted to occur at $ k\lambda_{D,c} \lesssim 0.15 $ with $ \gamma\omega_{p,c} \lesssim 5 \times 10^{-3} $, and increasing the beam temperature with respect to the core reduces the growth rate even further.

\pagebreak
\section{Numerical Simulation} \label{sec:simulation}


Time-dependent kinetic plasma simulations have been used to verify the linear theory and extend the study of the IAI into the non-linear
regime. Here, we present computations that adopt a Finite-Volume approach to discretize the Vlasov equation on a two-dimensional $x$-$v$-mesh:
Cell-averaged values of the $f_\alpha$ are stepped forward in time using a third order Runge-Kutta scheme \citep{SHU1988439},
where the numerical flux densities $vf_\alpha$ in the $x$-direction and $q_\alpha E f_\alpha / m_\alpha$ in the $v$-direction are averaged at cell
faces using one-by-one dimensional conservative parabolic reconstructions of $f_\alpha$ from either side,
resulting in a scheme of second order in $x$ and $v$.
Boundary conditions are periodic along $x$ and zero-flux along $v$, so that the total particle number in the domain is conserved
for each component, up to the resetting of potential negative undershoots of the $f_\alpha$ to zero.
Although core and beam protons consist of the same type of particles, we use separate phase spaces for all three components,
which allows us to follow the beam evolution in detail despite its relatively small density.
The domain length $L_x$ along $x$, resolved with $N_x = 200$ cells, differs between simulation runs, as specified below.
Velocity spaces are resolved with $N_v = 151$ cells that cover velocity intervals of size $L_{v,p} = 8 v_{th, c}$ for core and beam protons, and $L_{v,e} = 6 v_{th, e}$ for electrons,
centered around their individual drift velocities $V_\alpha$.

The electric field, located at cell faces along $x$, is simply integrated from eq. (\ref{eq:gauss}) with the condition $\int_0^{L_x} E(x)\, dx = 0$,
i.e., zero external electric field, in addition to $E(x=0) = E(x=L_x)$ which reflects periodicity.
The integration time step size is $\Delta t = 1 / (2500 \, \omega_{p, c})$.
Particle distribution functions are initialized as drifting Maxwellians, as described in Sec. \ref{sec:intro}. However, in order to break
the symmetry and excite the fundamental mode with wave number $k_1 = 2\pi / L_x$, the beam drift velocity $V_b$ in the exponential of eq. (\ref{eq:maxwellian})
is replaced by the slightly $x$-dependent function $V_b + 0.01 \sin(k_1 x)$, while the initial distributions of core and electrons are not changed.
The respective simulation domain sizes of \( L_x = 60 \,\lambda_{D,c} \) and \( L_x = 100 \,\lambda_{D,c}\), with corresponding wave numbers \( k_1 \approx 0.1 /  \lambda_{D,c} \) and \( k_1 \approx 0.0626 /  \lambda_{D,c} \)
of the fundamental mode, are selected in order to operate in the region of maximum growth as suggested by linear theory (see Fig. \ref{fig:disprel}).

\begin{figure}[ht!]
\plotone{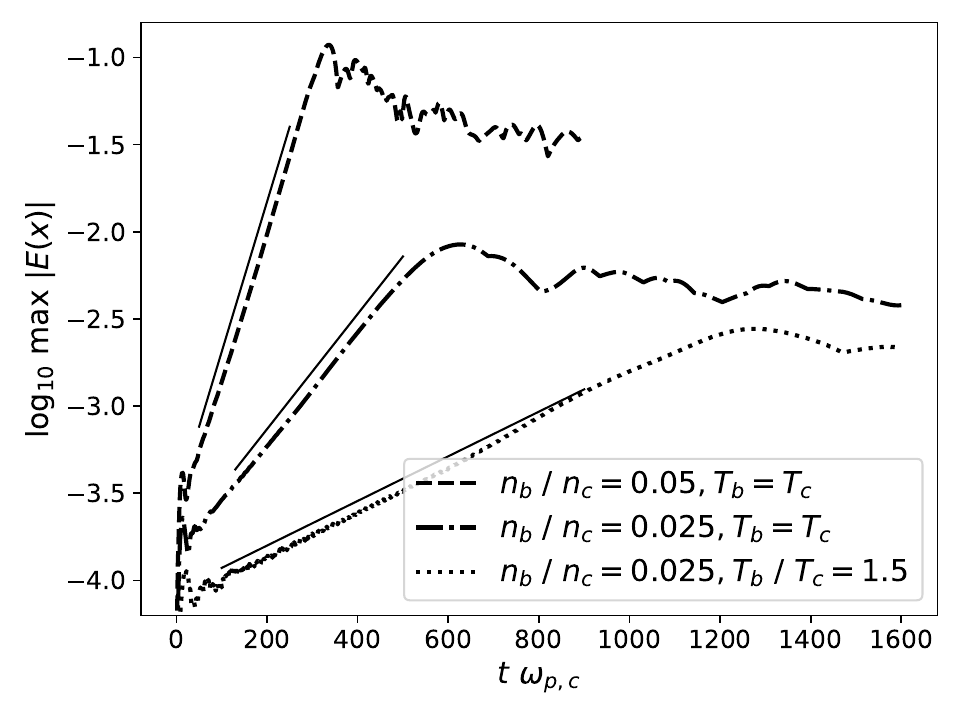}
\caption{
  Maximum value of the electric field, taken over the computational domain, vs. time for three simulations, runs
  with $T_e / T_c = 10$ and $V_D / V_{th,c} = 5$. Domain sizes are $L_x = 60 \lambda_{D,c}$  for
  runs with $T_b = T_c$, and $L_x = 100 \lambda_{D,c}$  for the case with $T_b / T_c = 1.5$.
  $E$ is normalized to $E_0 = m_c V^2_{th,c} / ( e \lambda_{D,c} )$.
  Straight lines fitted to the exponential phases
  correspond to $e$-based growth rates of $1.98\times 10^{-2}$, $7.6\times 10^{-3}$, and $2.95\times 10^{-3}\,\omega_{p,c}$, respectively.
  }
\label{fig:growth}
\end{figure}

\begin{figure}[ht!]
\plotone{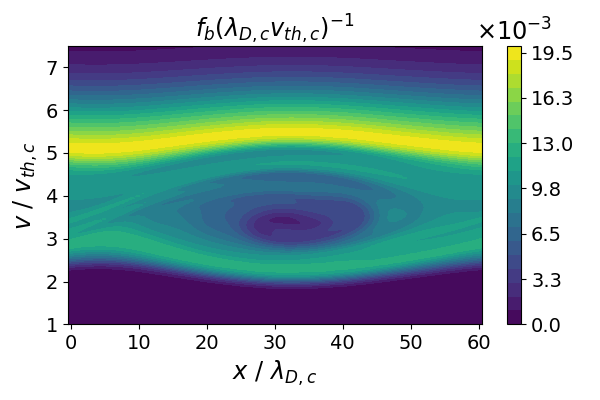}
\caption{
  Phase space density of the beam component from run $T_b = T_c$, $n_b / n_c = 0.05$ at time $t = 900 ~ \omega_{p,c}^{-1}$. This figure corresponds to the dashed curve of Fig. 3.}
\label{fig:phasespace1}
\end{figure}

\begin{figure}[ht!]
  \plotone{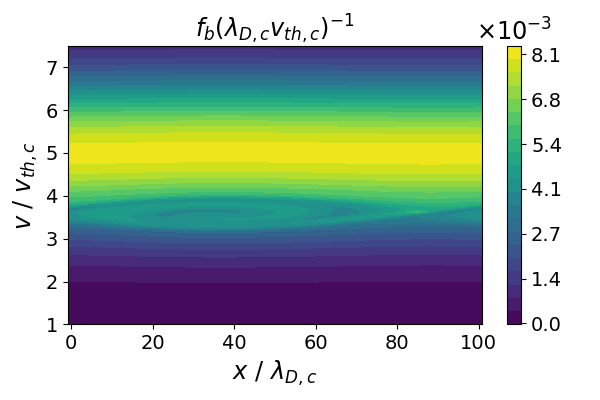}
\caption{
Phase space density of the beam component from run $T_b ~/~ T_c = 1.5$, $n_b / n_c = 0.025$ at time $t = 1500 ~ \omega_{p,c}^{-1}$. This figure corresponds to the dotted curve of Fig. 3.}
\label{fig:phasespace2}
\end{figure}

Fig. \ref{fig:growth} shows the temporal evolution of the three simulation runs with $n_b/n_c = 0.05$ and $0.025$,
as discussed in Sec. \ref{sec:theory}, where the maximum of the electric field magnitude inside the domain is used
as a time-dependent measure of the perturbation.
Growth rates measured during the exponential phases as inferred from Fig. \ref{fig:growth} are $1.98\times 10^{-2} $,
$7.6\times 10^{-3}$, and $2.95\times 10^{-3}\,\omega_{p,c}$, respectively.
These values are marked by stars in Fig. \ref{fig:disprel}, where comparison to the results from analytic theory
shows that the simulations yield slightly larger, but still compatible values, and the
effects of $n_b/n_c$ and $T_b/T_c$ on $\gamma$, as predicted by theory, are well confirmed.
Additional simulation runs, not shown here, with varying parameters and box sizes confirmed this picture together with the stability thresholds shown in Fig.~\ref{fig:stability}. The instability saturates into a phase with small overall changes, in which vortex patterns in the ion beam population are advected across the
domain (i.e., the frame of the plasma core), while they continue to wind up. Phase space plots from this stage are shown in Fig.
\ref{fig:phasespace1} for the case with $n_b/n_c = 0.05$, and in Fig. \ref{fig:phasespace2} for the weakest case with $n_b/n_c = 0.025$ and
temperature ratio $T_b/T_c  = 1.5$. During these stages, the ongoing filamentation in phase space is ultimately limited by the numerical mesh
resolution.
Moreover, Fig. 3 shows that a higher beam-to-core density ratio causes the instability to grow and reach the saturation phase faster than a lower ratio. Consequently, Figures 4 and 5 illustrate different times, i.e. the ends of the saturation phases.
Both Fig.~\ref{fig:growth} and the phase space plots indicate that the saturation level is clearly correlated with the strength of the instability in
the sense that faster growth comes with considerably larger distortions of the beam population and electric field fluctuations in the saturated
stage. Not shown here are the core ion populations: While the stronger instability ($n_b/n_c = 0.05$) produces a detectable distortion there too, with faint tongues of the core distribution reaching into the beam region,
the core remains almost unaffected in cases of weaker instability.

\begin{figure}[ht!]
  \plotone{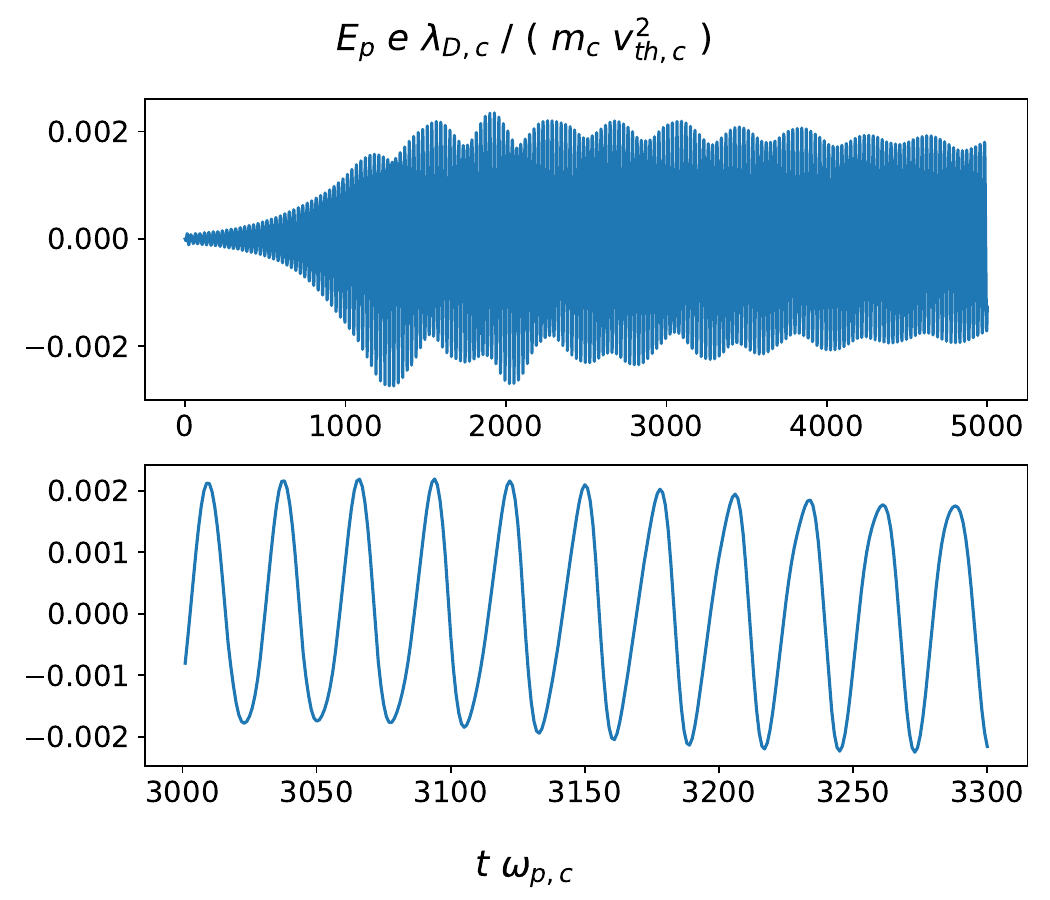}
\caption{
  The electric field was taken at the center of the domain for run  $T_b ~/~ T_c = 1.5$, $n_b / n_c = 0.025$.
  Top: Entire time simulation run. Bottom: Detail from time interval $t = (3000, 3300) ~ \omega_{p,c}^{-1} $.}
\label{fig:probe}
\end{figure}

\begin{figure}[ht!]
  \plotone{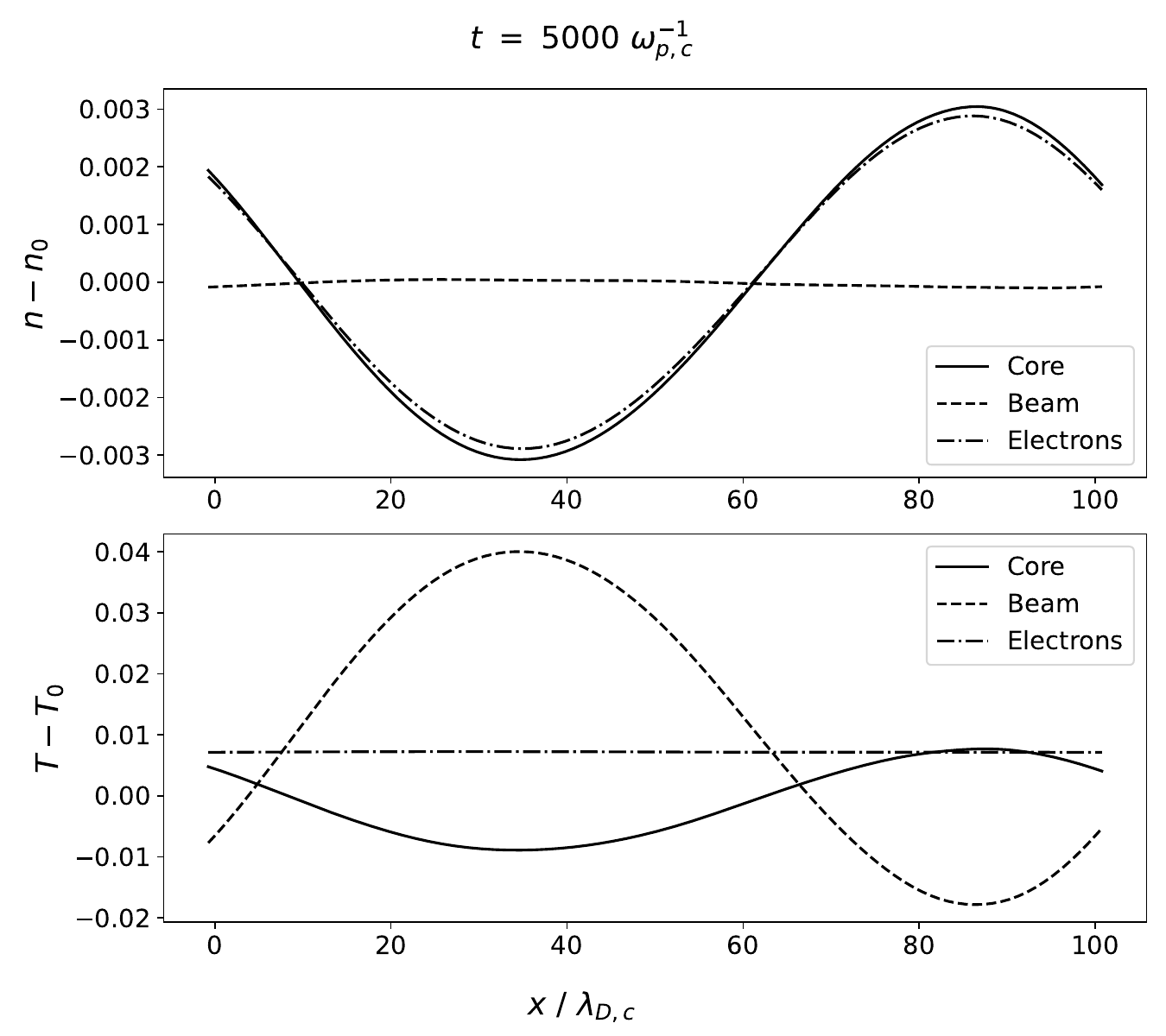}
\caption{
  Deviations of densities (top) and temperatures (bottom) from their respective initial values in
  run  $T_b ~/~ T_c = 1.5$, $n_b / n_c = 0.025$ at time $t = 5000 ~ \omega_{p,c}^{-1}$.}
\label{fig:heating}
\end{figure}

Fig.~\ref{fig:probe} shows $E$  at \( x = \frac{L_x}{2} \) for the case of weakest instability
over a longer simulation run time: while the first $\sim 1200$ plasma periods exhibit the onset with exponential growth, the saturated stage
continues for a long time to produce periodic sinusoidal signals, which result from the saturated islands being swept with nearly beam velocity across the probe position.
Consequently, the signal's period of $T \approx 28 \,\omega_{p,c}^{-1}$ at later times coincides with the ratio of wavelength $L_x = \frac{2\pi}{k_1} = 100\lambda_{D,c}$ and resonant velocity
$v_{res.}\approx 3.5\, V_{th,c}$. The burst's envelope corresponds to the exponential growth predicted by linear theory at early times and to a modulation due to slow global oscillation in the saturated stage. We note that the signal amplitude does not decay significantly over several $1000\,\omega_{p,c}^{-1}$.
Fig.~\ref{fig:heating} demonstrates once more the fairly small saturation level for the run with the weakest instability:
densities and temperatures of all three plasma components are only slightly modulated, where the largest relative effect of $\approx 10 \%$ with respect to initial values is in the beam density, reflecting the saturated island structures.
In particular, no substantial heating is observed in any of the components.

While all results presented here have been obtained with the mesh-based Vlasov simulation method described above, we have
carried out a number of additional simulations using the electrostatic particle-in-cell method described as ES1 in \citet{birdsallLangdon} and the electromagnetic particle-in-cell code (OSIRIS) \citep{Fonseca2002}. Both gave results that are in very good agreement with the ones presented here.

\section{Summary and Discussion} \label{sec:summary}
Motivated by recently reported observations with Parker Solar Probe (PSP)~\citep{Mozer2021a}, we have revisited key properties of the ion-ion acoustic instability in the regime of plasma parameters akin to those found in the solar wind. The study has been confined to a one-dimensional, purely electrostatic setup
representative of the plasma kinetic behavior parallel to an ambient magnetic field.
Using values close to those reported by
\citet{Mozer2021a} for the densities, temperatures, and drift velocities of the Maxwellian proton core, proton beam, and electrons, linear theory predicts a narrow parameter window conducive to the ion-ion-acoustic instability (IAI) onset. The three key parameters controlling IAI onset are I) the ratio of beam-to-core drift velocity to the core thermal speed, $V_D/ v_{th,c}$, II) the temperature ratio between electrons and core, $T_e/ T_c$, and III) the temperature ratio between beam and core, $T_b/ T_c$.
First, for the IAI to be unstable, the ratio of beam-to-core drift velocity to the core thermal speed has to lie within a relatively small interval around $ V_D / V_{th,c} \approx 5$, which matches the observed values quite well.
However, a second condition on the temperature ratio between electrons and ions suggests that, within our assumption of Maxwellian velocity distribution, values of
$ T_e / T_c \gtrsim 8-10 $ would be required for IAI onset, which is slightly larger than the reported $ T_e / T_c\approx 6.5$, see Table 1. Likewise, the beam temperature must not exceed the core temperature too much to avoid stabilization by resonant beam particles, as suggested by Figs. \ref{fig:stability} and \ref{fig:growth}. 
The fact that the observed parameter values are close to the stability threshold supports the idea of observed electric field bursts
being signatures of triggered IAI as put forward in \citet{Mozer2021a}. The crossings of the instability threshold could be imposed by varying
external conditions related to low-frequency activity found in the data.

Growth rates from linear theory are well-replicated in our grid-based Vlasov simulations that could follow single-mode growth of the instability
from small-amplitude initial perturbations up to saturation. In our simulations, we observe the formation of traveling vortices in the beam population, with associated signatures in densities and the electric field.

Reported PSP observations of fairly small density ratios of $n_b/n_c\approx 0.025$ \citep{Mozer2021a} lead us to
expect rather slow IAI growth and small saturation amplitudes.
Taking, for instance, the normalized growth rate to be $\gamma \approx 2\times 10^{-3}\,\omega_{p, c}^{-1}$ (cf. Fig. \ref{fig:disprel})
and ion-plasma frequencies of $f_{pi}\approx 9\,\mathrm{kHz}$ as reported by \citet{Mozer2021a}, the resulting growth time is estimated as
$\tau = 1/\gamma \approx 10\,\mathrm{ms}$, which appears to be in good agreement with Fig. 4 of \citet{Mozer2021a}.
Using the saturation amplitude and measured values of $T_{c,\|} \approx 7.7\,\mathrm{eV}$ (see Table 1) and $\lambda_{D,c}\approx 0.8\,\mathrm{m}$ to denormalize our simulated electric field,
our normalized electric field amplitude of $\approx 2\cdot 10^{-3}$ translates into values of $\approx 19\mathrm{mV/m}$.
This appears to be compatible with observations reported by \citet{Mozer2021a}, given the underlying uncertainties, strong sensitivities to parameter values, simplifications made on the theoretical side, and observational intricacy.

While reported observations include rapid decay of the high-frequency bursts, this feature is not reflected in our theory and simulation models,
where the phase space structures remain stable for a considerable time. However, our simulation domain at this point covers only one
wavelength of the perturbation and imposes periodicity through boundary conditions. One might speculate that in a more realistic
setting, the individual phase space vortices might undergo merging or similar breaks of symmetry that could lead to decay. At the same
time, the damping of the bursts as observed in PSP data could be associated with changing external conditions, in analogy to the triggering. These aspects are not accounted for in the present model and need further investigation.

An important open question concerns the temperature ratio $T_e/T_c$: IAI theory, supported by our numerical simulations,
would require values somehow larger ($\approx 8-10$) than the reported  ($\approx 6.5$) for the instability to exist. 
It is an object of ongoing investigation if this apparent discrepancy can be resolved by employing a more sophisticated model including anisotropies, oblique propagation with respect to the magnetic field, non-Maxwellian particle distributions, or coupling with other modes. These aspects will be addressed in future work.
On the other hand, a somewhat higher beam density than that reported from observations would shift the instability threshold towards smaller $T_e/T_c$ and could be an answer here.
What seems evident is that the IAI alone doesn't result in substantial electron heating, also because the free energy of
a very dilute proton beam will be much less than the electron's thermal energy and thus be unable to alter it significantly.
Other questions left unanswered by the present work are 
the issue of the triggering process and the decay mechanism of the instability, which are yet to be identified. 
Within this limit, however, it can be concluded that a quite simple one-dimensional electrostatic model seems to support the conjecture that the observed high-frequency bursts are
related to the ion-ion-acoustic instability.
 
\section*{Acknowledgement}
M. S. Afify thanks the Alexander-von-Humboldt Foundation (Bonn, Germany) for the research fellowship and its financial support. A.M. is supported by the Deutsche Forschungsgemeinschaft (German Science Foundation; DFG) project 497938371. K.S. and M.E.I. acknowledge support from DFG within the Collaborative Research Center SFB1491.

\bibliographystyle{aa}
\bibliography{iabursts}

\end{document}